\documentclass[sigconf]{acmart}
\usepackage{amsmath,amsfonts}
\usepackage{graphicx}
\usepackage{textcomp}
\usepackage{xcolor}
\usepackage{tikz}
\usetikzlibrary{positioning, shapes, shapes.geometric, shapes.symbols, shapes.arrows, shapes.multipart, shapes.callouts, shapes.misc}
\usepackage{listings}
\usepackage{url}
\usepackage[ruled,longend]{algorithm2e}
\usepackage{multirow}
\usepackage{textgreek}
\usepackage{wasysym}
\usepackage{subcaption}
\usepackage{footnotebackref}
\usepackage{tablefootnote}
\usepackage{threeparttable}
\usepackage{csquotes}
\usepackage{url}
\AtBeginDocument{%
  \providecommand\BibTeX{{%
    \normalfont B\kern-0.5em{\scshape i\kern-0.25em b}\kern-0.8em\TeX}}}

\copyrightyear{2022}
\acmYear{2022}
\setcopyright{acmcopyright}\acmConference[ICPC '22]{30th International Conference
on Program Comprehension}{May 16--17, 2022}{Virtual Event, USA}
\acmBooktitle{30th International Conference on Program Comprehension (ICPC '22),
May 16--17, 2022, Virtual Event, USA}
\acmPrice{15.00}
\acmDOI{10.1145/3524610.3529161}
\acmISBN{978-1-4503-9298-3/22/05}

\acmConference[ICPC 2022]{The 30th International Conference on Program Comprehension}{May 21–22, 2022}{Pittsburgh, PA, USA}
\begin{document}

\title{MSCCD: Grammar Pluggable Clone Detection Based on ANTLR Parser Generation}


% \affiliation{%

\author{Wenqing Zhu}
\affiliation{%
  \institution{Nagoya University}
  \city{Nagoya}
  \state{Aichi}
  \country{Japan}}
\email{zhuwqing1995@ertl.jp}

\author{Norihiro Yoshida}
\affiliation{%
  \institution{Nagoya University}
  \city{Nagoya}
  \state{Aichi}
  \country{Japan}}
\email{yoshida@ertl.jp}

\author{Toshihiro Kamiya}
\affiliation{%
  \institution{Shimane University}
  \city{Matsue}
  \state{Shimane}
  \country{Japan}
}
\email{kamiya@cis.shimane-u.ac.jp}

\author{Eunjong Choi }
\affiliation{%
 \institution{Kyoto Institute of Technology}
 \city{Kyoto}
 \state{Kyoto}
 \country{Japan}}
 \email{echoi@kit.ac.jp}

\author{Hiroaki Takada}
\affiliation{%
  \institution{Nagoya University}
  \city{Nagoya}
  \state{Aichi}
  \country{Japan}}
  \email{hiro@ertl.jp}

\renewcommand{\shortauthors}{Zhu and Yoshida, et al.}

\setlength{\belowcaptionskip}{-2pt} % !!! its OK? !!!

\begin{abstract}
 For various reasons, programming languages continue to multiply and evolve. It has become necessary to have a multilingual clone detection tool that can easily expand supported programming languages and detect various code clones is needed. However, research on multilingual code clone detection has not received sufficient attention. In this study, we propose MSCCD (Multilingual Syntactic Code Clone Detector), a grammar pluggable code clone detection tool that uses a parser generator to generate a code block extractor for the target language. The extractor then extracts the semantic code blocks from a parse tree. MSCCD can detect Type-3 clones at various granularities. We evaluated MSCCD's language extensibility by applying MSCCD to 20 modern languages. Sixteen languages were perfectly supported, and the remaining four were provided with the same detection capabilities at the expense of execution time. We evaluated MSCCD's recall by using BigCloneEval and conducted a manual experiment to evaluate precision. MSCCD achieved equivalent detection performance equivalent to state-of-the-art tools. 
\end{abstract}



\maketitle

\section{Introduction}

\begin{figure*}
  \graphicspath{{./img/}}
  \includegraphics[width=\textwidth]{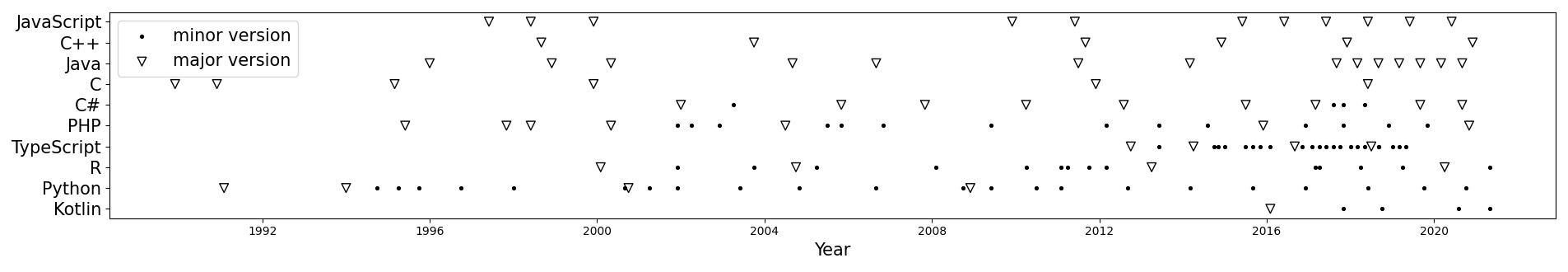}
  \caption{Release Log of Popular Languages}
  \label{fig:languageVersion}
\end{figure*}

Programming languages (hereinafter referred to as "languages") are advancing rapidly, and various languages are being developed and used for different purposes. Even within the same language, the syntax is frequently updated. 
For example, Typescript has been updated at least 26 times (see Figure \ref{fig:languageVersion}), including minor updates, since its first release in 2012. This is a blessing for practitioners who are free to choose the latest language for their purposes.

Code clone detection \cite{ccfinder,Sajnani16,NICAD}, a successful application of program analysis, is required to deal with a wide variety of languages and grammar definitions by practitioners \cite{Semura2017, Choi2015}. Code clone researchers frequently receive requests for clone detection tools that support new languages and grammatical definitions in their industry-academia collaboration activities. 

\textbf{However, it is unrealistic to support a wide variety of languages while frequently adapting to the changes in grammatical definitions  \cite{Semura2017}.} \textsf{NiCAD} \cite{NICAD} is a widely-used code clone detection tool that allows a user to specify the analysis method for each language. However, it is difficult for software developers who do not have much knowledge of program analysis to describe the analysis method.
Semura et al. developed a token-based clone detection tool, namely \textsf{CCFinderSW}  \cite{Semura2017}, which employs a lexical analysis mechanism to allow users to flexibly change the grammar of comments, identifier names, and keywords according to the target language. However, because \textsf{CCFinderSW} only supports lexical analysis changes, it cannot detect Type-3 clones \cite{Bellon} (see Section \ref{sec3}) that contain syntactic differences.

As a practical tool that supports a wide range of languages while flexibly responding to frequently changing grammar definitions, we propose \textsf{MSCCD} (\textbf{M}ultilingual \textbf{S}yntactic \textbf{C}ode \textbf{C}lone \textbf{D}etector), a Type-3 code clone detection tool that allows users to input ANTLR grammar definition files.
\textbf{The ability to input the grammar definition files of ANTLR, a widely used parser generator, is a practical design choice for developing a code clone detection tool that can respond quickly to frequently changing grammars}.
Because the grammars-v4 repository\footnote{https://github.com/antlr/grammars-v4} of ANTLR grammars has more than 150 grammar definition files and over 6000 commits since 2012, \textsf{MSCCD}, which can input ANTLR grammar files, can handle frequent grammar changes.

Once a user provides an ANTLR grammar file and target programs that follow the grammar, \textsf{MSCCD} detects Type-3 clones from the target programs based on the grammar.
By allowing the user to input the grammar definition file of ANTLR, \textsf{MSCCD} can be applied to programs written in many languages that have ANTLR grammar definition files.
\textsf{MSCCD} first generates a parser that extracts the token bags (i.e., collections of elements with duplicates of keywords, identifiers, and literals) from the program according to the grammar definition file specified by the user. It then uses a parser to generate the token bags \cite{Sajnani16} and detects similar subsequences to identify Type-3 clones that contain syntactic differences.

We investigated the language extensibility of \textsf{MSCCD} to programs written in 21 widely used languages (see Table \ref{fig:languages }). We applied \textsf{MSCCD} to programs written in each of the 21 languages included in the Rosetta Code,\footnote{\url{http://rosettacode.org/wiki/Rosetta_Code}} found that \textsf{MSCCD} can generate token bags for all the 20 languages whose grammar definition file exists in the grammars-v4 repository. In addition, we investigated the recall of \textsf{MSCCD} using a representative benchmark, BigCloneBench \cite{BCE, BCB} and found that the recall of \textsf{MSCCD} is comparable to that of \textsf{SourcererCC}  \cite{Sajnani16}, a state-of-the-art code clone detection tool (see Table \ref{Table:Recall_Precision}). Then, we evaluated the precision of \textsf{MSCCD} for the source code included in BigCloneBench using the same procedure applied in extant \textsf{SourcererCC} research  \cite{Sajnani16}. The results showed that \textsf{MSCCD} is slightly more precise than \textsf{SourcererCC} (see Table \ref{Table:Recall_Precision}).
Furthermore, \textsf{MSCCD} could complete the detection on a 100 MLOC source code collection in approximately 6 h (see Table \ref{Table:scalability}).

  The main contributions of this study are as follows: 
  \begin{itemize}
    \item We provide a tool, \textsf{MSCCD}, which detects Type-3 clones from a target program according to its grammar when the target programs and an ANTLR grammar definition file are given. To the best of our knowledge, \textsf{MSCCD} is the first Type-3 clone detection tool that can be used with a grammar definition file.

    \item Evaluation experiments show that \textsf{MSCCD} supports most of the widely used languages and is competitive with the state-of-the-art Type-3 clone detection tools in terms of quantitative measures, such as precision, recall, and execution speed.

    \item \textsf{MSCCD} and its experimental data are available on the Internet\footnote{\url{https://doi.org/10.5281/zenodo.5886550}}; this enables other researchers to reproduce the evaluation experiments.

  \end{itemize}

    The rest of this paper is organized as follows. Section \ref{sec2} describes the motivations for this research, including language diversity and release frequency. Section \ref{sec3} describes the important concepts and definitions. Section \ref{sec4} introduces the proposed approach (i.e., token bag generation using a parse tree (PT)) and the implementation of \textsf{MSCCD} in detail. Section \ref{sec5} describes various experiments conducted to answer the three research questions. Section \ref{sec6} introduces the threats to validation. Section \ref{sec7} presents the related work. Finally, Section \ref{sec8} concludes the paper and discusses our future plans.

\section{Motivation}
\label{sec2}
As programming languages evolve, code clone detection tools must keep pace. Figure \ref{fig:languageVersion} shows the release frequency of 10 popular languages (referring to the PopularitY of Programming Language Index ranking in September 2021\footnote{\url{https://pypl.github.io/PYPL.html}}). A triangle indicates a major update, and a dot indicates a minor update\footnote{For languages using the semantic versioning scheme (a system to manage version numbers semantically) \cite{SemanticVersion}, the major and minor versions are included. For other languages, larger versions equivalent to the major versions are included. }.
Nearly all the languages are updated regularly, and several languages appeared within a decade of each other. Each release is likely to have introduced lexical or syntactic changes to the grammar, and code clone detection tools must be updated as these changes occur. In most cases, developers need to modify the source program to support these updates, making it challenging to keep most of the existing clone detection tools up to date.

\begin{figure*}
  \graphicspath{{./img/}}
      \includegraphics[width=0.97\textwidth]{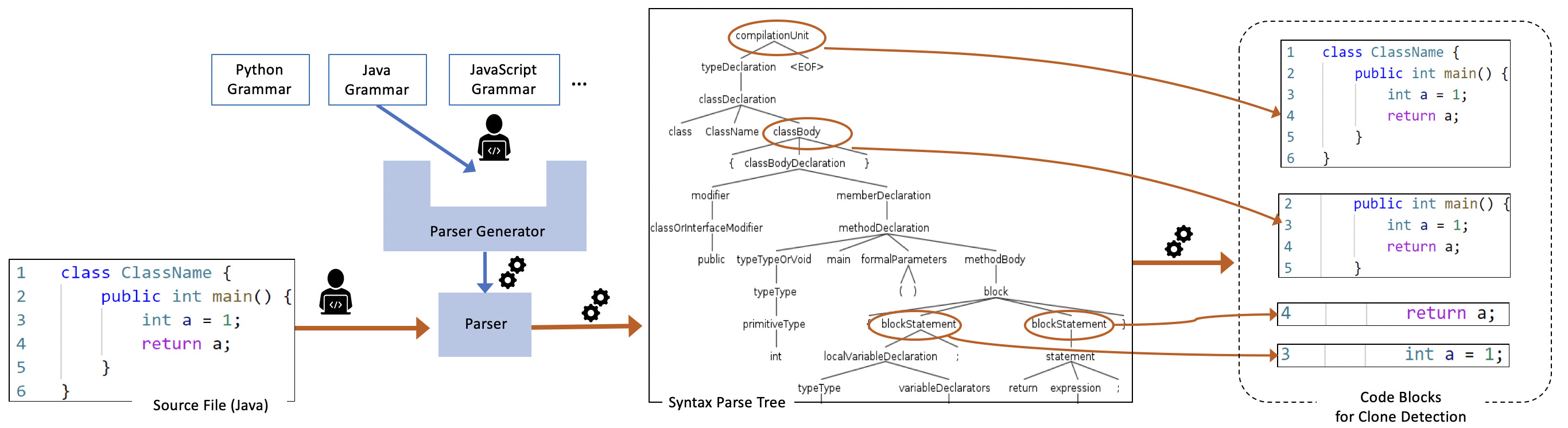}
  \caption{Multilingual Code Block Partition by Parse Tree}
  \label{fig:mainIdea}
\end{figure*}

Despite the large number of languages used in software development and the fact that these languages are regularly updated, the number of languages supported by most code clone detection tools is still limited. A survey paper on code clone detection research from 2013 to 2018 listed 13 tools  \cite{review2019}, of which only one had a language extension mechanism. 
The other 12 tools only support a limited number of languages, including Java and \textit{C/C++}. Research on code clone detection tends to be biased toward popular languages for which source code is plentiful and readily available.
To the best of our knowledge, there is no large-scale benchmark that evaluates the recall of code clone detection in other languages, as BigCloneEval  \cite{BCE} only supports Java.
Therefore, it is still difficult to evaluate the recall of code clone detection for other languages.

Among the existing tools, \textsf{CCFinderSW}  \cite{CCFinderSW} provides an extension mechanism to handle additional languages. This mechanism works by converting grammar definitions into regular expressions, targeting comments, string literals, and keywords. Regular expressions cannot express arbitrary context-free sentences; hence, \textsf{CCFinderSW} cannot support languages such as Lua. \textsf{CCFinderSW} only covers Type-2 clones, which is insufficient for many tasks. 
MSCCD can be used for many languages (probably more than 150) in the "grammars-v4" repository, and it can fully support new or updated grammars by simply reusing the ANTLRv4 grammars, which are actively developed by the "grammars-v4" community, in a drop-in manner.
Additionally, some tools emphasize easy expandability, such as \textsf{SourcererCC}  \cite{Sajnani16}. However, we found that sometimes these tools do not obtain correct results. For example, we used \textsf{SourcererCC}'s\footnote{https://github.com/Mondego/SourcererCC} block-level to tokenize a Java source file\footnote{\raggedright \url{https://github.com/tensorflow/Java/blob/daeb257/tensorflow-core/tensorflow-core-api/src/gen/annotations/org/tensorflow/op/DtypesOps.Java}}, but it failed to extract the correct blocks of some functions (lines 78--81) because the function parameters were split into several lines. Therefore, we are skeptical about the actual language extensibility of these tools.
\textsf{NiCAD} \cite{NICAD}, a widely-used code clone detection tool, allows the user to specify the analysis method for each language, but it is difficult for software developers who do not have much knowledge of program analysis to describe the analysis method.

Another possible solution to the rapidly changing language problem is to leverage an intermediate language such as the well-specified LLVM IR\footnote{\url{http://llvm.org}}. However, this solution cannot be used when the lexical and syntactic analyses of the program are required (e.g., syntactic clone detection  \cite{jiang2007deckard,baxter1998cloneast, Nguyen2009}). Additionally, there are several languages for which no conversion tools to LLVM IR-type have been developed or for which a proper conversion is labor-intensive (e.g., dynamically typed languages, such as Python).

Based on these observations, this work aims to develop a tool that detects Type-3 clones from a target program according to the corresponding grammar, given the target programs and an ANTLR grammar definition files are given.

\section{Terminology}
\label{sec3}
This paper uses the following definitions \cite{Roy09,roy2007survey}:

\noindent
\textbf{Token Bag}: A bag (i.e., a collection of elements with duplicates) of keywords, identifiers, and literals.

\noindent
\textbf{Granularity Value}: A non-negative integer indicating the level of the granularity of the code segment. Bigger granularity values correspond to the finer granularity.

\noindent
\textbf{Code Segment}: A section of continuous lines of code which is defined by the quaternion $(l,s,e,g)$, with the source file $l$, start line $s$, stop line $e$, and granularity value $g$.

\noindent
\textbf{Code Block}: A code block is a code segment whose sentences are grouped by one grammar rule.  

\noindent
\textbf{Composition}: In this paper, a composition is defined as a code block corresponding to at least one grammar rule. Classes, condition statements, loop statements, or functions can be a composition. This item is mainly used to evaluate the ability of MSCCD to generate token bags in Section \ref{sec5}.

\noindent
\textbf{Clone Pair}: A pair of similar code segments. 

\noindent
\textbf{Clone Type}: Code clones can be classified into four types:
\begin{itemize}
  \item \textbf{Type-1 (T1)}: Identical code segments, except for the differences in white-space, layout, and comments.
  \item \textbf{Type-2 (T2)}: Identical code segments, except for the differences in identifier names and literal values, in addition to the T1 clone differences.
  \item \textbf{Type-3 (T3)}: Syntactically similar code segments that differ at the statement level. The segments have statements added, modified, and/or removed with respect to each other, in addition to the T1 and T2 clone differences.
  \item \textbf{Type-4 (T4)}: Syntactically dissimilar code segments that implement the same functionality.
\end{itemize}

\section{Proposed tool}
\label{sec4}

The following subsections introduce the main idea and implementation of \textsf{MSCCD}.

\subsection{Main Idea: Code Block Partition by PT}
\label{sec:4.1}
Most code clone detection tools aim to not only detect clones between source files but also seek to partition them into code blocks \cite{Roy09}. This is a significant challenge for multilingual detection tools. Because the accurate division of code blocks requires syntax analysis and no syntax analyzer is suitable for multiple languages, replacing the syntax analyzer for the existing tools also requires source-code-level redevelopment. As shown in Figure \ref{fig:mainIdea}, the main idea is that \textbf{every subtree in a parse tree (PT) presents a semantic code block}. A PT is an ordered tree representing the syntactic structure according to grammar. It is generated via syntax analysis, presenting how production is applied to replace non-terminals. Thus, each subtree in a PT represents the production of the grammar, wherein the root node represents the left side of the production, and all child nodes of the root node represent the right side of the production. Correspondingly, all leaf nodes from the subtree present terminals, a token, an operator, or other lexical units. The source file can be divided into several blocks by handling these lexical units. Besides, a parser for the target language can be easily generated by using a parser generator. 

\begin{figure}
  \lstset{
numbers=left,
frame=lines,
language=Java,
basicstyle=\small,
xleftmargin=2em,
framexleftmargin=1.5em,
breaklines=true
}

  \begin{lstlisting}
block          : '{' blockStatements? '}';
blockStatements: blockStatement+ ;
blockStatement : localVariableDeclarationStatement
	       | classDeclaration
	       | statement ;
  \end{lstlisting}
  \caption{A Part of the Java Grammar}
  \label{fig:JavaBlockGrammar}
\end{figure}

Notably, the PT generated by a general-purpose parser is not suitable for code clone detection. On the one hand, a PT contains redundant nodes. 
For example, Figure \ref{fig:JavaBlockGrammar} shows an example of the Java 8 grammar, which defines code blocks. For the derivation of the production in line 2, if the non-terminal $blockStatement$ is matched only one time, the two subtrees (the root node of one is the non-terminal $blockStatements$, and the other is the non-terminal $blockStatement$) will contain the same lexical units. In other words, the two subtrees correspond to completely overlapping code blocks in the source file.
This phenomenon is common in all languages. On the other hand, not all nodes represent a code block that can be regarded as a semantic code block. For example, many tiny subtrees may correspond to a part smaller than a statement. These parts are meaningless for code clone detection.

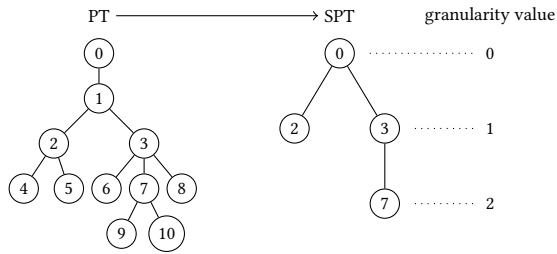
\begin{figure}
  \begin{tikzpicture}
    \tikzset{
      data/.style = {
      shape= tape,
      minimum size = 15pt,
      draw = black,
      align= center
    },
    treenode/.style = {
      shape=circle,
      inner sep = 2pt,
      minimum size = 2pt,
      draw = black,
      align= center
    },
    noneNode/.style = {
      draw = white,
      shape=circle,
      inner sep = 2pt,
      minimum size = 2pt,
      align= center
    },
    mrk/.style = {
      align= center
    }
    % ,
    % box/.style = {
    %   rectangle, 
    %   rounded corners, 
    %   inner sep=10pt, 
    %   inner ysep=20pt
    % }
    }
    \footnotesize

    % \node[data](grammar) at (-0.25, 1){Grammar};
    % \node[data](source code) at (0 , 0) {Source Code};
    \node[noneNode](a00) at (0, 0) {};

    % PT
    \node[treenode](a00) at (1.8, 1.5) {0};
    \node[treenode](a10) at (1.8, 0.9) {1};
    \node[treenode](a20) at (1.2, 0.3) {2};
    \node[treenode](a21) at (2.4, 0.3) {3};
    \node[treenode](a30) at (0.8, -0.3) {4};
    \node[treenode](a31) at (1.4, -0.3) {5};
    \node[treenode](a32) at (1.9, -0.3) {6};
    \node[treenode](a33) at (2.4, -0.3) {7};
    \node[treenode](a34) at (2.9, -0.3) {8};
    \node[treenode](a40) at (2.1, -0.9) {9};
    \node[treenode](a41) at (2.7, -0.9) {10};
    
    \draw[-] (a00) -- (a10);
    \draw[-] (a10) -- (a20);
    \draw[-] (a10) -- (a21);
    \draw[-] (a20) -- (a30);
    \draw[-] (a20) -- (a31);
    \draw[-] (a21) -- (a32);
    \draw[-] (a21) -- (a33);
    \draw[-] (a21) -- (a34);
    \draw[-] (a33) -- (a40);
    \draw[-] (a33) -- (a41);
    
    % SPT
    \node[treenode](b00) at (5.0, 1.5) {0};
    \node[treenode](b10) at (4.4, 0.5) {2};
    \node[treenode](b11) at (5.6, 0.5) {3};
    \node[treenode](b20) at (5.6, -0.5) {7};

    \draw[-] (b00) -- (b10);
    \draw[-] (b00) -- (b11);
    \draw[-] (b11) -- (b20);

    %granularity
    \node[noneNode](s0) at (7, 1.5) {0};
    \node[noneNode](s1) at (7, 0.5) {1};
    \node[noneNode](s2) at (7, -0.5) {2};

    \draw[dotted] (5.4,1.5 ) -- (6.8,1.5 );
    \draw[dotted] (6,0.5 ) -- (6.8,0.5 );
    \draw[dotted] (6,-0.5 ) -- (6.8, -0.5 );

    \node[mrk](m1) at (1.8, 2) {PT};
    \node[mrk](m2) at (5.0, 2) {SPT};
    \node[mrk](m3) at (7, 2) {granularity value};

    \draw[->] (m1) -- (m2);
  \end{tikzpicture}
  \begin{tablenotes}
    \item *: The minimum tokens is set to 2.
  \end{tablenotes}
  \caption{ Simplification of a Parse Tree }
  \label{figure:blockExtraction}

\end{figure}

\begin{table}
  \caption{The Number of Extracted Code Blocks}
  \label{tab:codeBlockNum}
\centering
\renewcommand{\arraystretch}{1.2}
\begin{tabular}{l|l|l} 
\hline
id & strategy                   & extracted code blocks  \\ 
\hline
1 &  extract from PT    & $1\,525\,922$       \\ 
\hline
2 & extract from simplified SPT & $60\,541$                  \\ 
\hline
3 & id 2 with keyword filter             & $35\,556$                  \\
\hline
\end{tabular}

  \begin{tablenotes}
    \item *: The strategy 1 is not implemented in MSCCD. 
  \end{tablenotes}

\end{table}

\begin{algorithm}
  \caption{Parse Tree Simplification}
  \label{alg:simplify}
    \LinesNumbered
%   \SetAlgoLined
  \SetKw{merge}{merge}
  \SetKw{delete}{delete}
  \KwIn{ $T$ is a PT, each tree node contains an attribute $size$ representing its token number and an attribute $child$ containing child nodes; $mSize$ is the configured minimum size }
  \KwOut{ An SPT }

  \SetKwProg{Fn}{Function}{:}{end}
  \SetKwComment{Comment}{//}{}

  \Fn{ParseTreeSimplification{\upshape(}$T$, $mSize${\upshape)}}{ 
    \ForEach{{\upshape tree node} $n$ {\upshape in pre-order traversal of} $T$ }{
      \While{$n${\upshape .length} == $n${\upshape .child[0].length}}{
        \merge($n$, $n$.child[0])
      }
     
    \If{$n${\upshape .size} \textless\  $mSize$}{
      \ForEach{{\upshape child node} $cn$ {\upshape from} $n$}{
        \delete($cn$)
      }
    }
    }
    \Return{$T$}
  }

\end{algorithm}

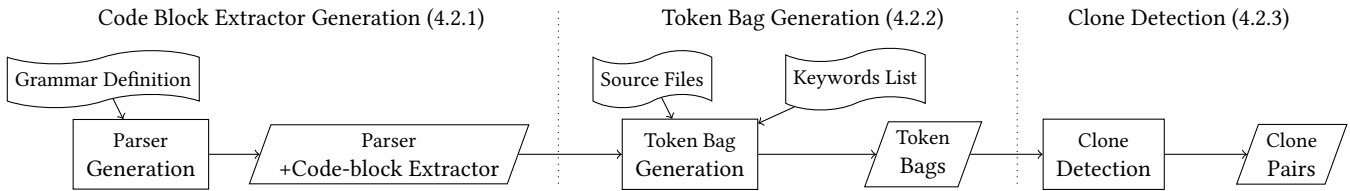
\begin{figure*}
\begin{tikzpicture}[node distance=10pt]
  \tikzset{
    input/.style = {
      shape= tape,
      minimum size = 10pt,
      draw = black,
      align= center
    },
    process/.style = {
      rectangle,
      inner sep = 5pt,
      minimum size = 10pt,
      draw = black,
      align= center
    },
    data/.style = {
      trapezium,
      trapezium left angle=70,
      trapezium right angle=-70,
      minimum size = 10pt,
      draw = black,
      align= center
    },
    mrk/.style = {
      align= center
    }
  }

%   \node[input](grammar) at (-0.3,0) {Grammar\\ Definition};
%   \node[process](ANTLR) at (-0.2,1.3) {Parser\\ Generation};
%   \node[data](parser) at (2.2,1.3){Parser};
%   \node[process](normalizerG) at (2.2,0) { Normalizer \\ Generation };
% 図を修正してみました。間違ってたら戻してください。???
  \node[input](grammar) at (-0.3,1.-0.2) {\small Grammar Definition};
  \node[process](ANTLR) at (0.2,-0.2) {\small Parser\\Generation};
  \node[data](parser) at (3.5,-0.2){\small Parser\\+Code-block Extractor};
%   \node[process](normalizerG) at (1.5,0) { Code-Block Extractor\\Generation };
  
%   \node[data](normalizer) at (5,0) {Code Block\\ Extractor };
  \node[input](keywords) at (9.7,0.8) {\small Keywords List};
  \node[input](Source) at (7,0.8){\small Source Files};
  \node[process](normalization) at (7.5,-0.2){\small Token Bag \\ Generation};
  \node[data](tokenBags) at (10.6,-0.2) {\small Token\\ Bags};
  \node[process](CloneDetection) at (13,-0.2){\small Clone\\ Detection};
  \node[data](Clone) at (15.5,-0.2) {\small Clone\\ Pairs};

  \draw[->] (grammar) -- (ANTLR);
  \draw[->] (ANTLR) -- (parser);
%   \draw[->] (parser) -- (normalizerG);
  \draw[->] (parser) -- (normalization); % ここも修正を入れた箇所です。???
  \draw[->] (keywords) -- (normalization);
%   \draw[->] (normalizerG) -- (normalizer);
  \draw[->] (Source) -- (normalization);
  \draw[->] (normalization) -- (tokenBags);
  \draw[->] (tokenBags) -- (CloneDetection);
  \draw[->] (CloneDetection) -- (Clone);
%   \draw[->] (normalizer) -- (normalization);

  \draw[dotted] (5.75, 1.7) -- (5.75,-0.6);
  \draw[dotted] (11.85,1.7) -- (11.85,-0.6);
  \node[mrk](m1) at (2.2, 1.6) { Code Block Extractor Generation (\ref{sec:4.2.1})};
  \node[mrk](m2) at (9, 1.6) { Token Bag Generation (\ref{sec:4.2.2})};
  \node[mrk](m3) at (14, 1.6) {Clone Detection (\ref{sec:4.2.3})};
\end{tikzpicture}
\caption{Overview of MSCCD}
\label{fig:stmProcess}
\end{figure*}

To reduce meaningless data, we propose simplifying the PT and generating a simplified PT (SPT). Algorithm \ref{alg:simplify} lists the steps required to simplify the PT. The input PT is traversed in the pre-order (line 2). The first step for each node, $n$, is to merge all child nodes containing the same lexical units as $n$ (lines 3--5). The merged nodes are not visited afterward. The second step is to check if node $n$ contains sufficient lexical units to meet the configured value (line 7). If node $n$ is not large enough, all child nodes of $n$ will be deleted to make $n$ a leaf node (lines 8--10). The deleted nodes will not be visited in the traversal afterward. Figure \ref{figure:blockExtraction} shows an overview of the simplification when setting $mSize$ to two.

Each node in the SPT corresponds to a subtree in the original PT. Thus, it corresponds to a code block. The depth of a node in the SPT is defined as its granularity value. The smaller the depth value, the coarser the granularity. Furthermore, nodes having the same depth value are more likely to represent the precise composition of the language. For example, in Java, nodes having a depth value of one represent classes, and those having a value of four represent functions.

Additionally, languages use keywords to define semantic code blocks, such as classes, methods, and loops. These code blocks are also targets for code clone detection. This class of nodes in the SPT has a feature that has a child node corresponding to a subtree in the PT that only contains keywords. This feature allows for further filtering of nonsensical code blocks. We call it a keyword filter. It is discussed in Subsection \ref{sec:4.2.2}.

To verify the effectiveness of PT simplification and the keyword filter, we conducted a preliminary experiment with the source code of Tensorflow-Java\footnote{https://github.com/tensorflow/Java}. It contains 2,158 files with 563,059 lines of code. We calculated the number of code blocks extracted using this method under various conditions. The results are summarized in Table \ref{tab:codeBlockNum}. When using the keyword filter, the number of target nodes was only approximately 2\% of the original non-leaf nodes. Additionally, depending on the detection task's needs, the range of block size, granularity values, and keywords can also be configured to limit further the number of code blocks involved in clone detection.

The proposed method has three merits. 
First, \textbf{ source files can be correctly divided into semantic blocks.} Because all subtrees in the PT correspond to a production in the grammar, the generated code blocks also contain the semantic information from that production. Additionally, via syntax analysis, the correctness of the code block partition can be ensured. 
Second, \textbf{Code blocks at various granularities can be generated by using this method.} In the PT, the node closer to the root node represents a greater granularity, such as a class or the whole file; the node farther from the root node represents a finer granularity, such as that of a statement. Thus, detectors can detect clones at multiple granularities or within a specific language component. 
Third, \textbf{high language extensibility can be ensured using a parser generator}. Theoretically, a parser generator can generate parsers for all context-free languages. The parsers generated by the same parser generator can be accessed using the same application programming interface (API). As a result, the supporting language can be changed without modifying the program.

\subsection{Overview of MSCCD}
\label{sec:4.2}
The overview of the proposed code clone detection tool (i.e., \textsf{MSCCD}) is summarized in Figure \ref{fig:stmProcess}. This depiction can be regarded in three phases: code block extractor generation (Section \ref{sec:4.2.1}), token bag generation (Section \ref{sec:4.2.2}), and clone detection (Section \ref{sec:4.2.3}). First, \textsf{MSCCD} generates a code block extractor for the target language using a parser generator. Second, the input source code is converted into token bags. Finally, MSCCD detects the code clones between the token bags. Algorithm \ref{Alg:TokenBagGeneration} lists the main steps of token bag generation in Section \ref{sec:4.2.2}.

\subsubsection{Code Block Extractor Generation}
\label{sec:4.2.1}

Here, \textsf{MSCCD} generates a code block extractor for the target language. For each language, \textsf{MSCCD} only needs a grammar definition file. Thus, \textsf{MSCCD} can change the supporting language or language version by 
changing only the grammar definition file, and there is no need to modify the program. Hence, \textsf{MSCCD} has excellent language extensibility. This only needs to be executed once for each language.

We chose ANLTR\footnote{https://www.antlr.org/}, which can create a parser for a language based on the grammar definition, including lexer and parser rules. For ANLTR, a language is defined by context-free grammar expressed using the extended Backus–Naur form (BNF). Therefore, grammar definition files for ANTLR can be easily created using the official grammar information. For languages that need unique treatments (e.g., the indents in Python), ANTLR allows programs to be embedded in the generated parser to solve the issue. Furthermore, there is a GitHub repository named "grammars-v4" \footnote{https://github.com/antlr/grammars-v4}, which is a collection of ANTLR grammars and handling programs with more than 150 languages. Users can easily obtain grammar definition files from this repository for a target language.

The code block extractor simplifies the PT generated by the parser using the algorithm introduced in Section \ref{sec:4.1} and identifies the code blocks.

\begin{algorithm}
  \caption{Token Bag Generation}
  \label{Alg:TokenBagGeneration}
  \LinesNumbered
  \KwIn{ $T$ is a SPT generated by Algorithm 1. $K$ is the keywords list.  }
  \KwOut{A Collection of token bags}
  
  \SetKwProg{Fn}{Function}{:}{end}
  \SetKwRepeat{Do}{do}{while}
  \SetKwComment{Comment}{//}{}
  
\Fn{TOKENBAGGENERATION  {\upshape(} $T$, $K$ {\upshape)}}{
    $TargetNodes$,$TokenBags$ = [];
    
    \eIf{$K$ == $\varnothing$ }{
        \ForEach{TreeNode $n$ in $T$}{
            $TargetNodes$.append($n$);
        }
    }{
        \ForEach{TreeNode $n$ in $T$}{
            \If{KeywordsFilter($n$, $K$) == True}{
                $TargetNodes$.append($n$);
            }
        }
    }
    \ForEach{ TreeNode $tn$ in $TargetNodes$ }{
        $TokenBags$.append( token bag created from $tn$ )
    }
    \Return{ $TokenBags$ };
  }
  
\Fn{KeywordsFilter {\upshape(} $n$, $K$ {\upshape)}}{
    \ForEach{ TreeNode $cn$ in $n.childs$}{
        $flag$ = 0;
        
        \ForEach{ Token $t$ in $cn$}{
            \eIf{ $t$.type == 'Keyword'}{
                $flag$ |= 1;
            }{
                $flag$ |= 3;
            }
        }
        \If{ $flag$ == 1}{
            \Return{ True};
        }
    }
    \Return{ False};
  }
\end{algorithm}

\subsubsection{Token Bag Generation}
\label{sec:4.2.2}

The source files from the input project are partitioned and converted into token bags during this phase. The input software project, $P$,  consists of several files, $F$: $P = \left \{ F_{0}, F_{1} , ... , F_{n}\right \}$. A file, $F$, can be represented by a set of code blocks, $SG$. \textsf{MSCCD} allows overlap between code blocks because they have different granularity values, and overlaps are bound to occur. \textsf{MSCCD} transforms these code blocks into token bags  \cite{zhang2010understanding}, which are defined as sets of tokens.

For each source file, MSCCD first generates an SPT using Algorithm \ref{alg:simplify}. Then MSCCD generates a token bag for each extracted code block. Algorithm \ref{Alg:TokenBagGeneration} lists the steps of token bag generation. The input includes an SPT and a keyword list. All the sub-trees corresponding to code blocks is saved in the list $TargetNodes$. If the keywords filter is not activated, all the nodes from SPT will be set as target nodes which presents a code block (line 3-7). Only the nodes that passed the keywords filter are set as target nodes (line 8-14). At last, MSCCD generates token bags for each target node by accessing the corresponding sub-tree in the original PT (line 18). 

The keyword filter can be activated by providing a non-empty keyword list. The input sub-tree can pass the keyword filter if at least one child only contains keywords (line 22-33). By using this filter, \textsf{MSCCD} retains more semantic code blocks to reduce the number of candidates. Additionally, this feature can be used to select specific parts of the language. For example, the user can detect code clones only between functions in Python by providing a keyword list that only contains the keyword "def".

\subsubsection{Clone Detection}
\label{sec:4.2.3}

Code clones are detected by finding similar pairs from the generated token bags. To detect clones, we adopted the definition and algorithm proposed by \textsf{SourcererCC}, which has good recall and scalability for syntactic code clone detection  \cite{Sajnani16}. It uses the ratio of the number of tokens shared by two token bags to the number of elements in the larger bags to present similarity. Two token bags, $B_{x}$ and $B_{y}$, are judged as clones if the similarity is greater than or equal to the threshold, $\theta $:
\begin{equation}
  \frac{\left  |B_{x}\cap B_{y}  \right |}{MAX\left ( \left|B_{x} \right|, \left| B_{y}\right| \right )} \geq \theta \\.
\end{equation}

The main challenge in this part is organizing the token bags during detecting clones to obtain better performance. As explained in Section \ref{sec:4.2.2}, a source file is presented by several token bags with several granularity values. Furthermore, there may be overlaps between token bags at different granularities. Theoretically, detecting clones between all the candidate pairs can achieve the highest recall. Correspondingly, this strategy has the highest operating overhead, significantly increasing execution time. Notably, the operating overhead increases due to the massive comparison times, and removing the overlapped clones from the report takes time. Overlapped clones are generated because the corresponding code segments may also be cloned, e.g., the corresponding sub-fragment of a T1 clone pair will still be a T1 code clone.

We chose to detect clones only between token bags with the same granularity value. In this strategy, when setting the biggest granularity value as $g_{max}$ and the number of token bags in granularity value $i$ as $N_{i}$, the time complexity of candidate comparison is reduced from $O\left ( \left ( \sum_{i=0}^{g_{max}} N_{i} \right  ) ^{2} \right) $ to $O\left ( \sum_{i=0}^{g_{max}} N_{i}^{2} \right )$. This is mainly because the functional consistency of code cloning makes it more likely to have the same granularity value, especially for inner-project clones. When the cloned code segment has a granularity value close to that of the original segment, there is a high probability that the clone will be detected due to the existence of the overlapped token bags. In other words, there will be a token bag containing the code segment at the corresponding granularity. Although the similarity of the code segments participating in the comparison is reduced, they can still be detected if higher than the threshold. Additionally, this method can be easily parallelized by using multiple processes to compare token bags in each granularity value.

For the overlapped clones, we choose only to report those of the smallest granularity value. The remaining overlapping clones will be filtered out after detection.

\section{Evaluation}
\label{sec5}
To evaluate the performance of \textsf{MSCCD}, we conducted experiments to answer the following research questions:

\begin{itemize}
  \item \textbf{RQ1}: How many languages can \textsf{MSCCD} support?
  \item \textbf{RQ2}: What is the precision/recall and the scalability of \textsf{MSCCD} code clone detection?
  \item \textbf{RQ3}: Are the code clones detected by \textsf{MSCCD} for each language appropriate for the purpose of software maintenance?
\end{itemize}

\subsection{RQ1: Language Extensibility}

\label{subsec:5.1} 

\textbf{Experimental Design}: The proposed method includes an optional keyword filter. When executing \textsf{MSCCD} without using the keyword filter, it can theoretically support all languages with ANTLRv4 grammar files. However, this strategy increases the execution time because of a higher overlap of candidates. On the other hand, some code blocks might be filtered out incorrectly when using the keyword filter, thus reducing recall. This experiment evaluates the number of compositions \textsf{MSCCD} supports with or without using the keyword filter.  

We determined four items for each language: class, function, branch, and loop. These items are the most basic and standard parts of languages. For each item, all corresponding grammar is included within the scope of the test targets. For example, all loop items, including but not limited to the while loop, for loop, and do-while loop, are test objects. We regard function items as the most relevant to show the extensibility of the target language among the four items. The other three items reflect whether \textsf{MSCCD} can detect code clones at more granularities. However, it should be noted that \textsf{MSCCD} does not merely support these four items. The other language components can also be accurately extracted depending on the grammar. If all corresponding code blocks are accurately extracted, including the corresponding code block, lacking any irrelevant code segments, the item is passed (circle mark; \CIRCLE). It is regarded as a failure if a target code block fails to be generated correctly without a syntax analysis error.

\noindent
\textbf{Target Language and Input Data:} We evaluated the most popular 21 languages according to the PYPL ranking. For each language, we randomly selected five files with more than ten lines from the Rosetta Code.\footnote{\url{http://www.rosettacode.org/wiki/Rosetta_Code}}. Rosetta Code is a site that collects solutions to more than 1,138 tasks in various languages. It adopts basic methods and does not use a development framework; hence, the four items will appear in the code more commonly. If the test targets are randomly selected from open source software, the chosen targets are sometimes unable to obtain some items because these compositions may not be used in the source program.

\noindent
\textbf{Tool Configuration:} In this experiment, we set the minimum tokens as two, which can fully expose the language
component extraction ability of \textsf{MSCCD}. Moreover, the complete keyword list (the keyword list which contains all the keywords according to official information) is provided.

\begin{table}
  \vspace{-3mm}
  \small
  \caption{Language Extensibility to the top 21 most popular languages in PYPL}
  \label{fig:languages }
\setlength\tabcolsep{3.5pt} % default value: 6pt
\begin{tabular}{lcccc|lcccc}
\hline
Lang.        & F. & Cl. & Cd. & Loop & Lang. & F. & Cl. & Cd. & Loop \\ \hline
Python      &\CIRCLE \CIRCLE         &\CIRCLE \CIRCLE         & \CIRCLE \CIRCLE        &\CIRCLE \CIRCLE         & 
Kotlin      &\CIRCLE \CIRCLE         &\CIRCLE \CIRCLE         &\CIRCLE \CIRCLE         &\CIRCLE \CIRCLE         \\ 
Java        &\CIRCLE \CIRCLE         &\CIRCLE \CIRCLE         &\CIRCLE \CIRCLE         &\CIRCLE \CIRCLE         & 
Matlab      &\LEFTCIRCLE \LEFTCIRCLE &\sun                    &\LEFTCIRCLE \LEFTCIRCLE &\LEFTCIRCLE \LEFTCIRCLE \\ 
JavaScript & \CIRCLE \CIRCLE        &\sun                    & \CIRCLE \CIRCLE        &\CIRCLE \CIRCLE          & 
Go          &\CIRCLE \CIRCLE         &\sun                    &\CIRCLE \CIRCLE         & \CIRCLE \CIRCLE        \\ 
C\#         &\CIRCLE \CIRCLE         &\CIRCLE \CIRCLE         &\CIRCLE \CIRCLE         &\CIRCLE \CIRCLE         & 
VBA         & \CIRCLE \CIRCLE        &\sun                    & \CIRCLE \CIRCLE        & \CIRCLE \CIRCLE        \\
PHP         & \CIRCLE \CIRCLE        &\CIRCLE \CIRCLE         &\CIRCLE \CIRCLE         &\CIRCLE \CIRCLE         & 
Rust        &\CIRCLE \CIRCLE         &\CIRCLE \CIRCLE         &\CIRCLE \CIRCLE         &\CIRCLE \CIRCLE         \\
C           &\CIRCLE \CIRCLE         &\sun                    &\CIRCLE \CIRCLE         &\CIRCLE \CIRCLE         & 
Ruby        &\LEFTCIRCLE \hexstar    &\LEFTCIRCLE \hexstar    &\LEFTCIRCLE \LEFTCIRCLE &\LEFTCIRCLE \LEFTCIRCLE \\
C++         &\CIRCLE \CIRCLE         &\CIRCLE \hexstar        &\CIRCLE \CIRCLE         & \CIRCLE \CIRCLE        & 
Scala       &\LEFTCIRCLE \LEFTCIRCLE &\LEFTCIRCLE \LEFTCIRCLE &\LEFTCIRCLE \LEFTCIRCLE &\LEFTCIRCLE \LEFTCIRCLE \\
R           &\LEFTCIRCLE \LEFTCIRCLE &\sun                    &\LEFTCIRCLE \LEFTCIRCLE &\LEFTCIRCLE \LEFTCIRCLE & 
Ada         &\diameter               &\diameter               &\diameter               &\diameter               \\
TypeScript &\CIRCLE \CIRCLE         &\CIRCLE \CIRCLE         &\CIRCLE \CIRCLE         &\CIRCLE \CIRCLE          & 
VB          &\CIRCLE \CIRCLE         &\sun                    &\CIRCLE \CIRCLE         &\CIRCLE \CIRCLE         \\
Swift       &\CIRCLE \CIRCLE         &\CIRCLE \CIRCLE         &\CIRCLE \CIRCLE         &\CIRCLE \CIRCLE         & 
Dart        &\CIRCLE \CIRCLE         &\CIRCLE \hexstar        &\CIRCLE \CIRCLE         &\CIRCLE \CIRCLE         \\
Objective-C &\CIRCLE \hexstar        &\sun                    &\CIRCLE \CIRCLE         &\CIRCLE \CIRCLE         & 
            &                        &                        &                        &                        \\
\hline
\end{tabular}

\footnotesize{\flushleft{\noindent
1: \CIRCLE: Positive   \hexstar: Negative  \sun: Not applicable \LEFTCIRCLE: Positive, compilation error exists  \diameter: Grammar definition does not exist\noindent\\
2: Left side: result without keyword filter. Right side: result with keyword filter. \noindent\\
3: Lang.: Language. F.: Function, Cl.: Class, Cd.: Condition.\\
}}
\end{table}

\noindent
\textbf{Result}: Table \ref{fig:languages } lists the experiment results. Among the 21 target languages, 20 are available in \textsf{MSCCD}. Moreover, all the components of these 20 languages can be supported when not using the keyword filter. Among the 20 languages, functions of 18 are successfully extracted except for Objective-C and Ruby. We manually checked the productions of functions in Ruby's grammar. We found that the root node of all the corresponded subtrees of the function composition can not contain a child node that only contains leaf nodes of keywords. Therefore, these subtrees in Ruby are filtered out by the keywords filter. 
For the same reason, only a part of the functions can be supported in Objective-C. All languages support the condition of the other three items, and the loop is supported by 19, except for Go. The reason for failing to extract the loop in Go is the same as that of the functions in Objective-C. For class, 11 passed the experiment among the 13 languages. The overall experiment results indicate that \textsf{MSCCD} has relatively high language extensibility. 

Notably, the experiments of four languages (left circle mark; \LEFTCIRCLE) had syntax analysis errors and failed to generate a correct PT for some source files. Because the purpose of this experiment was to evaluate how many languages can be supported when the PT is generated, we manually checked the grammar of these languages to check the fact that these compositions can be extracted or not.

\noindent
\textbf{Comparison with CCFinderSW}: Table \ref{fig:languages } implies that \textsf{MSCCD} can be used for many languages in the ``grammars-v4'' repository. With state-of-the-art tools, \textsf{CCFinderSW} \cite{CCFinderSW} has a close level of language extensibility. However, CCFinderSW's approach cannot support some languages e.g., Lua, making its language extensibility lower than that of \textsf{MSCCD}. One of the necessary steps for CCFinderSW to support a language is to convert the grammar rules into regular expressions. This conversion is not always possible since regular grammar is a subset of ANTLR's context-free grammar. Since MSCCD performs the syntactic analysis using ANTLR, such an issue does not arise. In addition, MSCCD can fully support new or updated grammars by simply reusing the ANTLRv4 grammars in a drop-in manner.

\noindent\fbox{
  \parbox{0.45\textwidth}{
The answer to RQ1:
When using the keyword filter, \textsf{MSCCD} can provide function-level support for 18 of 20 available languages. Furthermore, when the keyword filter is inactivated, \textsf{MSCCD} can support all 20 tested languages. 
  }
}

\subsection{RQ2: General Performance}
\label{subsec:5.2}
We believe that because the similarity calculation results were the same, the detection performances of \textbf{MSCCD} remain unchanged regardless of the language. The experiments were conducted using Java. To compare the existing syntactic code clone detection tools, we followed the experimental methods published in \textsf{SourcererCC}  \cite{Sajnani16} and used the framework and dataset provided therein. For all the experiments discussed in this subsection, we configured \textsf{MSCCD} for a minimum clone size of 50 tokens and a similarity threshold of 70\%. A complete list of keywords is provided. The evaluation results of other compared tools were taken from the published work  \cite{CCAligner}.

% \subsubsection{\textbf{Recall}}
\hyphenation{Big-CloneEval}
\hyphenation{Big-CloneBench}

\noindent
\textbf{Recall}: We measured the recall of \textsf{MSCCD} by using Big-CloneEval \cite{BCE}, which provides clone detection tool evaluations based on BigCloneBench \cite{BCB}. BigCloneEval reports the recall of each type of code clone, which is used to measure the detection ability of syntactic code clones. In BigCloneEval, T3 clones were divided into Very-Strongly Type-3 (VST3), Strongly Type-3 (ST3), Moderately Type-3 (MT3), and Weakly Type-3 (WT3) according to the similarity \cite{BCE}. We set the clone matcher to a minimum size of six lines and 50 tokens for comparing the existing results \cite{CCAligner}.

Table \ref{Table:Recall_Precision} shows the result of recall measurements. Similar to these tools, \textsf{MSCCD} also has a near-perfect recall on T1 and T2 clones. For the recall of the three T3 clones, \textsf{MSCCD} ranks third. Notably, the comparison with \textsf{SourcererCC} is interesting. The reason for this is explained as follows. The two tools use basically the same similarity calculation method and clone detection algorithm. In theory, the recall based on BigCloneBench should be the same. However, \textsf{MSCCD} recall on ST3 and MT3 is higher than that of \textsf{SourcererCC}. We believe that this is because \textsf{MSCCD} creates token bags in multi-granularity. When performing a complete detection, the same code segment is inspected at multiple granularities, improving the detection ability. To prove this argument, we set a test group for \textsf{MSCCD} containing only reported clones in granularity values 0 and 4 (file-level and function-level in Java). In this group, the recall of ST3 and MT3 dropped slightly and was closer to that of SourcererCC. The detection results in granularity values 0 and 4 only account for approximately 44\% of the total. \textsf{MSCCD} reported a large number of clones at other granularities. 

% \subsubsection{\textbf{Precision}}
\noindent
\textbf{Precision}:
We measured the precision of \textsf{MSCCD} using the same random sample test as \cite{Sajnani16}. We randomly selected 400 clone pairs detected by \textsf{MSCCD} in the BigCloneEval experiment and equally distributed them to five judges to determine the correctness of each clone pair. The clone pairs were marked as "unknown" for judging complex situations. To compare the state-of-the-art tools, we set two test groups. One group contained pairs selected from all the results. This group represents the average precision of MSCCD. The other group only contained reported pairs in granularity values 0 and 4. This group is more convincing to compare with the other tools because the granularity is closer to traditional tools. 

Table \ref{Table:Recall_Precision} lists the results. For the first group containing results for all the granularities, the precision of \textsf{MSCCD} was calculated from the resulting 368 pairs of true positives and 32 pairs of false positives (including eight pairs marked as difficult to determine). \textsf{MSCCD} had the precision at 92\%. For the group that only contains results in granularity values 0 and 4, \textsf{MSCCD} also had a near precision of 91\%.

\begin{table}[]
\caption{Recall and Precision Measurements}
\label{Table:Recall_Precision}
\begin{tabular}{c|ccccc|c}
\hline
\multirow{2}{*}{Tool} & \multicolumn{5}{c|}{Recall}  & \multirow{2}{*}{Precision} \\ \cline{2-6}
                    & T1  & T2  & VST3 & ST3 & MT3 &                            \\ \hline
MSCCD                 & 100 & 98  & 93   & 63  & 7   & 92                         \\
MSCCD*                & 100 & 98  & 93   & 61  & 6   & 91                         \\
SourcererCC           & 100 & 98  & 93   & 61  & 5   & 83                         \\
CCAligner             & 100 & 99  & 97   & 70  & 10  & 80                         \\
CCFinderX             & 100 & 93  & 62   & 15  & 1   & 72                         \\
Deckard               & 60  & 58  & 62   & 31  & 12  & 60                         \\
NiCad                 & 100 & 100 & 100  & 95  & 1   & 56                         \\
iClones               & 100 & 82  & 82   & 24  & 0   & 91                         \\ \hline
\end{tabular}
\begin{tablenotes}
\item[1] *: only contains clone pairs in granularity value 0 and 4
\end{tablenotes}
\end{table}

\begin{table}

  \caption{Execution Time}
  \label{Table:scalability}
  \resizebox{0.48\textwidth}{!}{
  \renewcommand{\arraystretch}{1.2}
  \begin{tabular}{c|clllll}
    \hline
    LOC  & 1K & 10K & 100K & 1M       & 10M        & 100M \\ \hline
    Time & 1 sec & 4 sec  & 17 sec  & 3 min 13 sec & 1 hr 14 min 33 sec &  6 hr 6 min 52 sec    \\ \hline
    % Time* &    &     &     &    &   &      \\ \hline
    % Function &  &   &   &  &  &      \\ \hline
  \end{tabular}
  }
\end{table}

\noindent
\textbf{Execution Time}:
To evaluate the performance of different data sizes, we generated test files by randomly selecting files from IJaDataset\cite{IJaDataset}. We used the Linux command \enquote{wc} to measure lines of codes. The experiments were executed on a quad-core CPU, and the maximum heap memory size of the Java Virtual Machine was set to 12GB. The time to generate the parser and code block extractor was less than 5 s and only 
needed to be executed once. We also set two test groups to measure execution time when the keyword filter is activated or when it is not. The result is listed in Table \ref{Table:scalability}. \textsf{MSCCD} has good scalability and can complete the 100-MLOC-level clone detection task in slightly more than 6 hr. 

\noindent\fbox{
  \parbox{0.45\textwidth}{
The answer to RQ2:
\textsf{MSCCD} has a level of recall and precision equivalent to state-of-the-art tools and can complete detection on repositories of up to 100MLOC.
  }
}

\subsection{RQ3: Language Features of the Detected Clones}
\label{subsec:5.3}

While some code fragments of code clones correspond to the logic, others are difficult to use for maintenance, such as a sequence of import or constant declarations. In this section, we present an experimental evaluation of whether \textsf{MSCCD} could detect code clones of code fragments that represent logic in a wide range of minimum token numbers, which is a code clone detection parameter. 

In the experiment, we selected five repositories from GitHub for nine target languages by order of stars. We believe that the high-star repositories are more representative. \textsf{MSCCD} was executed twice for each repository when configuring the minimum token number to 20 and 50. Table \ref{Table:tested data} lists an overview of the experiment, including size, number of extracted token bags, number of clones detected, and the ratio of keywords. 

\begin{figure*}
  \centering
  
  \begin{subfigure}{\textwidth}
  
  \includegraphics[width=0.19\textwidth]{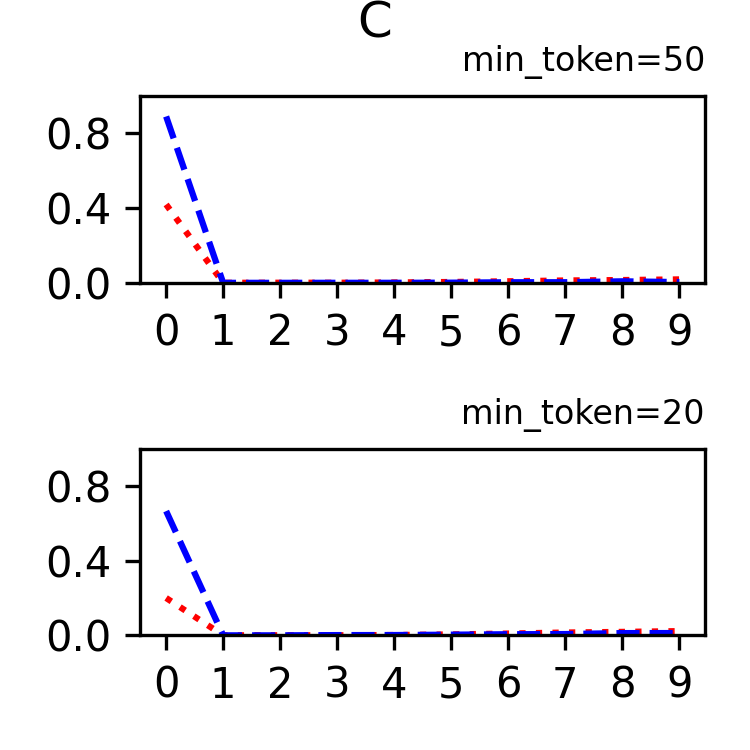}
  \includegraphics[width=0.19\textwidth]{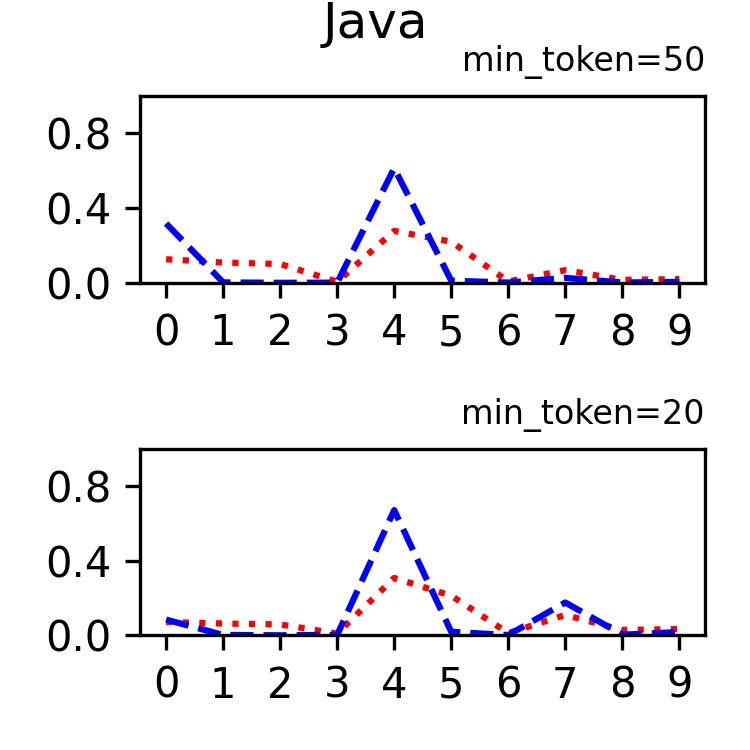}
  \includegraphics[width=0.19\textwidth]{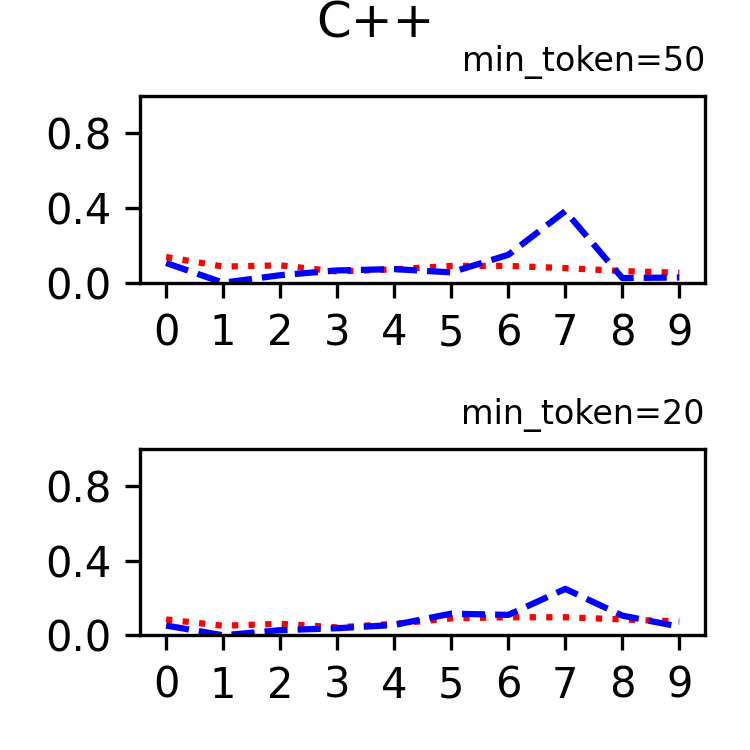}
  \includegraphics[width=0.19\textwidth]{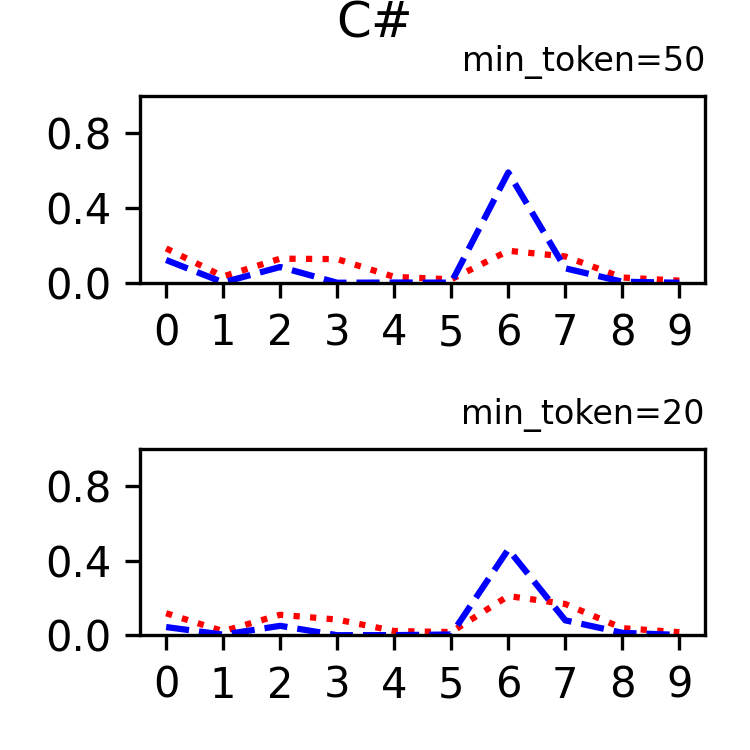}
  \includegraphics[width=0.19\textwidth]{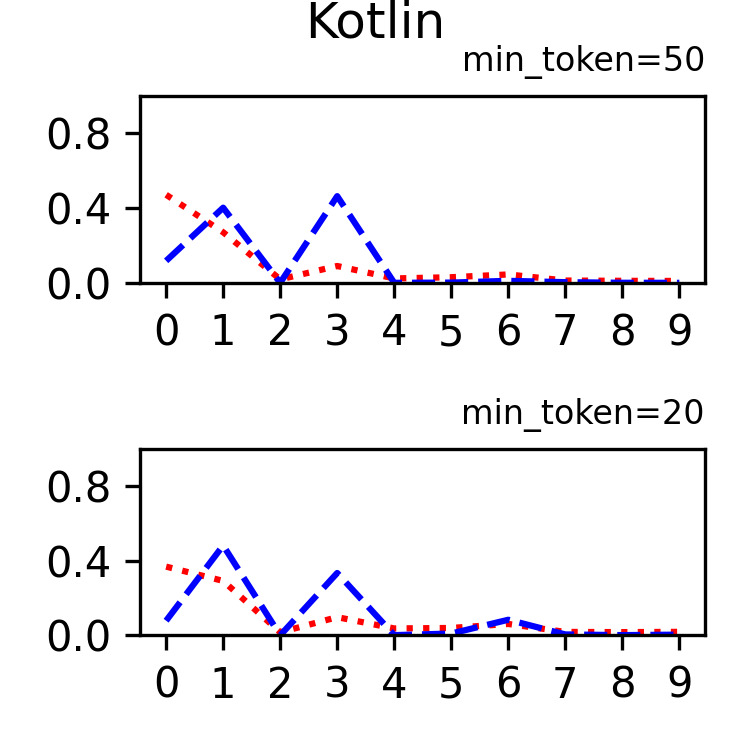}
  
  \end{subfigure}
  \begin{subfigure}{\textwidth}
  
    \includegraphics[width=0.19\textwidth]{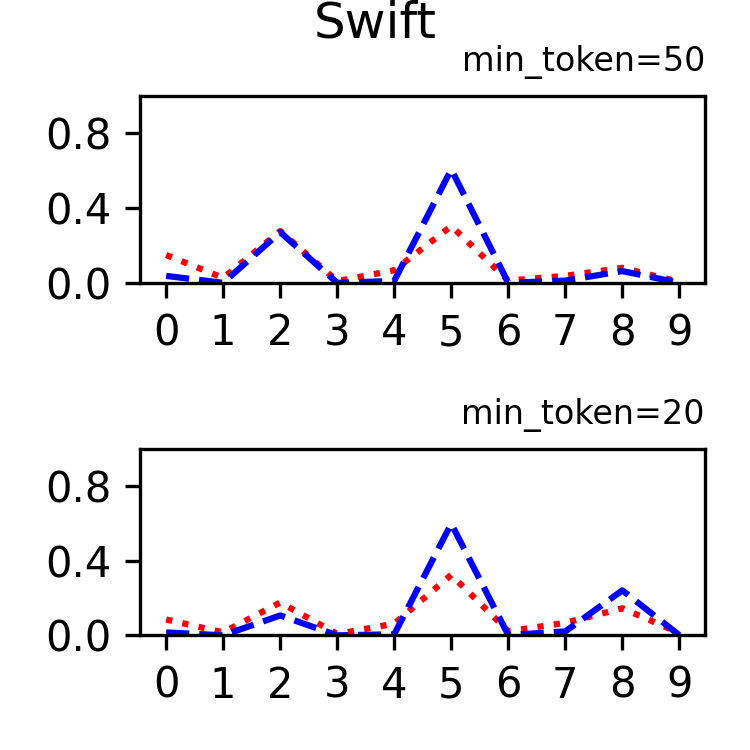}
    \includegraphics[width=0.19\textwidth]{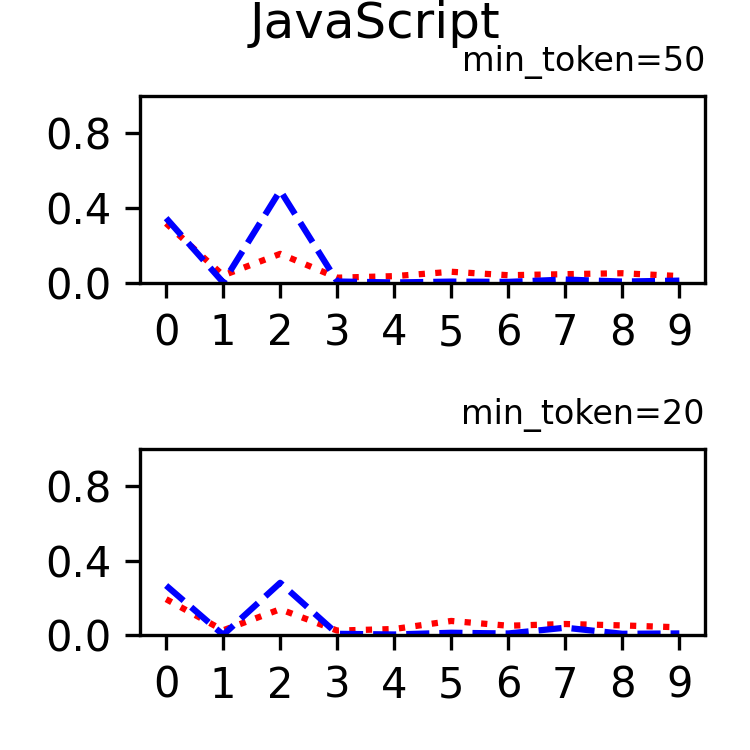}
  \includegraphics[width=0.19\textwidth]{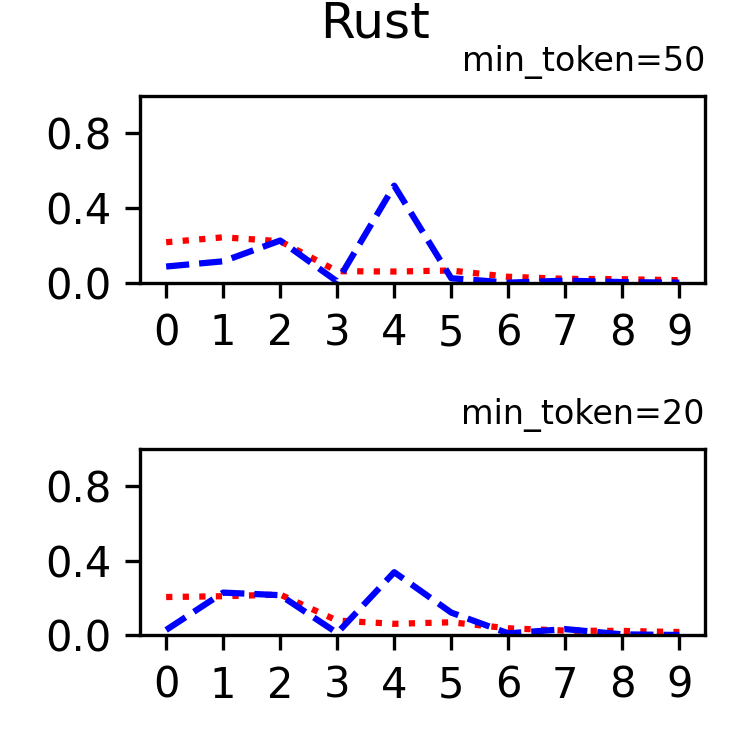}
  \includegraphics[width=0.19\textwidth]{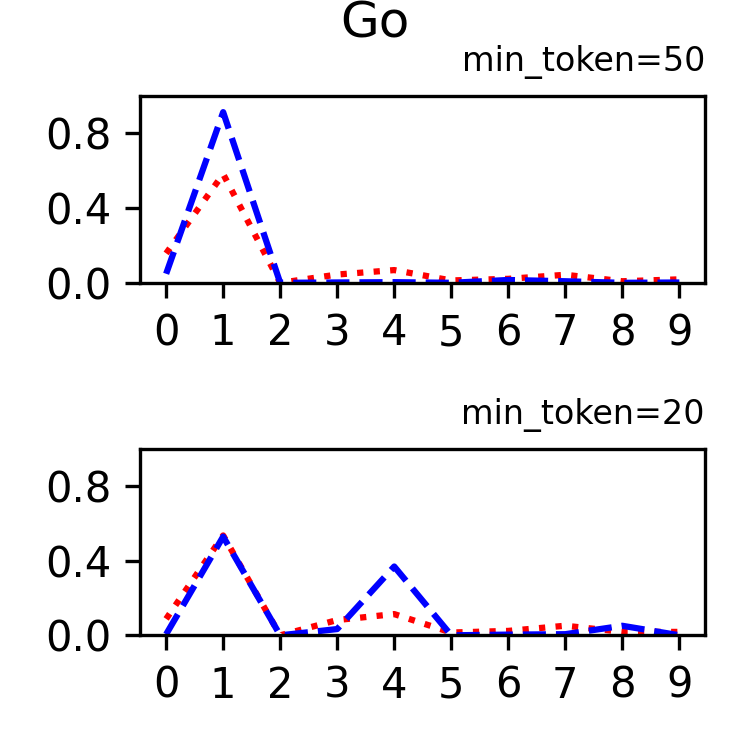}
  \hspace*{0.2in}
  \includegraphics[width=0.19\textwidth]{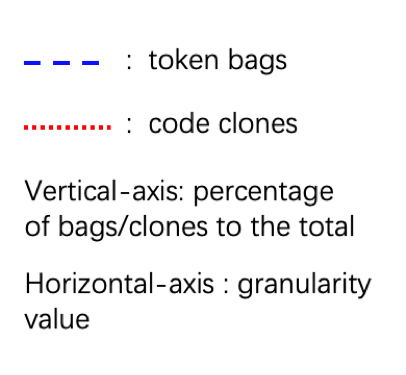}
%   \makebox[0pt][l]{\hspace*{0.7cm}\raisebox{2.1ex}{\includegraphics[width=0.1\textwidth]{img/legend_new.png}}} 
%   \includegraphics[width=0.1\textwidth, trim = 1cm,0,0,0]{img/legend.png}
  \end{subfigure}
  
  \caption{Distributions of Token Bags and Clones}
  \label{fig:dictributions}
\end{figure*}

The language compositions that can be described at each granularity are different depending on the grammar. Therefore, we investigated how the granularity of the generated token bag and the detected clone changes when the min token is changed for each language. We drew the distribution of all the extracted token bags and all the token bags that were detected as clones along with each granularity value into a line graph (Figure \ref{fig:dictributions}). The x-axis of each graph represents the granularity value, and the y-axis represents the token bags and the reported clones corresponding to that granularity.

\begin{table}
  \caption{Overview of Detections in 9 languages}
  \label{Table:tested data}
  \small
  \resizebox{0.48\textwidth}{!}{
  \renewcommand{\arraystretch}{1.2}
  \begin{tabular}{l|c|c|c|cc|c}
\hline
\multirow{2}{*}{Language} & \multirow{2}{*}{MLoc} & Tokens                    & Token Bags                    & \multicolumn{2}{c|}{Detected Clone} & Keywords \\
                          &                      & (M)                        & (K)                           & mt = 20    & mt = 50   & Ratio    \\ \hline
C                         & 21.09                  & 48.80                     & 3405.84                      & 10191            & 7148             & 15.44\%  \\
Java                      & 4.62                   & 13.20                     & 391.61                       & 504966           & 96004            & 37.68\%  \\
C++                       & 3.14                   & 9.22                      & 126.18                       & 55317            & 16823            & 47.89\%  \\
C\#                       & 2.10                   & 4.35                      & 117.98                       & 100178           & 22333            & 62.40\%  \\
Kotlin                    & 2.33                   & 7.58                      & 169.51                       & 31805            & 16554            & 39.53\%  \\
Swift                     & 0.28                   & 0.65                      & 18.82                        & 59672            & 20121            & 49.94\%  \\
JavaScript                & 0.95                   & 2.57                      & 39.32                        & 22230            & 10439            & 14.27\%  \\
Rust                      & 1.65                   & 3.91                      & 90.90                        & 6419             & 2040             & 38.48\%  \\
Go                        & 7.42                   & 22.20                     & 35.52                        & 9342416          & 1019142          & 10.30\%  \\ \hline
\end{tabular}
  }
    \begin{tablenotes}
     \item[1] mt: min token. (M): mega. (K): kilo. 
   \end{tablenotes}
  \end{table}

% \begin{minipage}{0.25\textwidth}
%   \fbox{The distribution of code clones in different languages is also different. For some languages (C\#, Go, and so on), more clones can be detected by adjusting the tool's configuration, such as the smaller minimum tokens or finer granularity. At the same time, the precision may decrease slightly due to the increase in the number of candidates.}
% \end{minipage}
\begin{table}[]
  \caption{Frequently Occurring Compositions in Figure \ref{fig:dictributions}}
  \label{table:composition}
    \resizebox{0.48\textwidth}{!}{
    \setlength\tabcolsep{3.5pt} % default value: 6pt
    \begin{tabular}{lcl|lcl|lcl}
    \hline
    Lang.                   & GV & Comp.        & Lang.                  & GV & Comp.        & Lang.                   & GV & Comp. \\ \hline \hline
    Java                    & 4  & function     & \multirow{3}{*}{Swift} & 2  & class        & \multirow{2}{*}{Go}    & 1  & function     \\ \cline{1-3} 
    C++                     & 7  & branch, loop &                        & 5  & function     &                        & 4  & branch, loop \\ \cline{1-3} \cline{7-9}
    C\#                     & 6  & function     &                        & 8  & branch       &  JavaScript          & 2  & function       \\ \hline
    \multirow{2}{*}{Kotlin} & 1  & class        & \multirow{2}{*}{Rust}  & 2  & function     & \multicolumn{3}{c}{GV: Granularity Value} \\
                            & 3  & function     &                        & 4  & branch       & \multicolumn{3}{c}{Comp.: Composition} \\ \cline{1-6}
    \end{tabular}
    }
  \end{table}

The first point of concern is the number and position of the heap in the polyline representing the clone (blue polylines). The heap in the blue polyline indicates that many clones were gathered at this granularity. Table \ref{table:composition} lists frequently occurring compositions at the granularity where there is a heap in Figure \ref{fig:dictributions}. We believe that function-level clones are more efficient for software maintenance. The clones in branches and loops are usually error handling, exception capture, and high-frequency code (such as traversal), which are relatively insignificant for software maintenance.

When the minimum number of tokens was changed from 50 to 20, some languages did not change the heap position very much, while others changed significantly. The small clones detected are often part of the larger clones in the languages whose heap did not change. Many small clones exist in the language whose heap changed that are not part of the large clones. So many tiny clones were reported. For example, in Go, the position of the clone curve heap changed significantly when the minimum number of tokens was 20. It occurred because Go's error handling generates an abundance of clones between branches. By contrast, for languages such as Java and JavaScript, the position of the heap does not change, but the height drops slightly. It showed few clones with smaller granularity values in this language, and most were produced in larger units. Tiny clones may reduce the effectiveness of the result set for maintenance but may be adequate for other tasks. Detection tools can be configured according to requirements.

Unlike other languages, the code clone of the C language is gathered at the granularity value of 0 whenever the minimum number of tokens is set to 20 or 50. However, this is not a feature of C language syntax but is related to application scenarios. The experimental object of the C language in this experiment includes the source code of the Linux Kernel. Many file-level clones exist because they need only individual replacements to adapt to different devices. The Linux Kernel project is significantly larger than the other tested projects, so the characteristics of the project overshadow the C language's characteristics.

\noindent\fbox{
  \parbox{0.45\textwidth}{
The answer to RQ3:
For most of the languages, \textsf{MSCCD} detected code clones such as functions, branches, and loops, which would be of interest for maintenance tasks when the parameter min token ranged from 20 to 50. For some languages, different min tokens may result in different kinds of clones being detected, which might be tuned by monitoring granularity.
  }
}

\section{Threats to Validity}
\label{sec6}

One internal threat is that we did not report the precision of different languages. We designed the experiments in Section 5.2 based on the following assumptions: Because \textsf{MSCCD} converts code blocks written in context-free languages into token bags, there will be no significant difference in the precision between different languages. The result shows that \textsf{MSCCD} has the same precision as clone detection tools for a specific language because \textsf{MSCCD} adopts the same detection method as the latest Type-3 clone detection tool (i.e., SourcererCC) after converting code blocks to token bags. Besides, \textsf{MSCCD} does not tune for any particular language. Making benchmarks by ourselves for this study may lack appropriateness and objectivity, so we abandoned the solution of reporting accuracy for each language separately and instead used the classical benchmark BigCloneBench for indirect evaluation.

We also did not report the recall of other languages. Firstly, we believe that the performance of the detection method is consistent in most programming languages. Secondly, there is a lack of recall evaluation benchmarks (e.g., BigCloneBench \cite{BCB}), making it impossible to provide recall for all supported languages. Besides, BigCloneBench only contains code clones at the function level. That means the recall of clones with below granularities is not measured. As mentioned in Section \ref{sec5}, the number of clones reported at granularity values 0 and 4 is only about 44\% of the total. From this, we are optimistic about the recall of the finer granularity. We will seek newer evaluation methods to measure that in future work.

Another internal threat is that the data tested in the experiment discussed in Section \ref{subsec:5.3} may not be sufficient. We selected the repositories based on the number of stars on GitHub. Because high-star open-source software (OSS) is used more frequently and has a more demonstrable effect, we believe that the high-star repositories can represent language situations. However, there may be significant differences in project sizes, which brings about the problem of uneven weighting. We will continue to expand the scope of the experiment in the future.

An external threat exists regarding the language extensibility of \textsf{MSCCD}. \textsf{MSCCD} tokenizes code based on parsers generated by ANTLRv4, and there may be parsing errors. If an error occurs at an early stage, for example, during lexical analysis or at a higher position than target subtrees in the PT, \textsf{MSCCD} will not output the expected results. Users must confirm that the used grammar definition file is executable in ANTLRv4 and that it matches the version with the target files.

\section{Related work}
\label{sec7}
Since the 1990s, many code clone detection tools and methods have been proposed. Based on the similarity analysis approach, these tools can mainly be categorized in to four classes: \cite{Roy09} textual \cite{NICAD, ducasse1999language}, lexical \cite{ccfinder,cp-miner,basit2007efficient}, syntactic \cite{baxter1998cloneast,jiang2007deckard,kontogiannis1996pattern}, and semantic \cite{liu2006gplag,gabel2008scalable}. The following keywords are commonly found in the research hits:

\noindent
\textbf{T3 clone detection}: \textsf{NiCAD}  \cite{NICAD}, \textsf{iClones}  \cite{gode2009incremental}, \textsf{Deckard}  \cite{jiang2007deckard}, \\
\textsf{CCAligner}  \cite{CCAligner}, and \textsf{SourcererCC}  \cite{Sajnani16} are some well known detection tools that detect code clones up to T3. In the experiment, \textsf{MSCCD} had the same level of recall as these tools. Moreover, \textsf{Oreo}  \cite{saini2018oreo} and \textsf{CloneWorks}  \cite{svajlenko2017fast} also have good performance in T3 code clone detection.

\noindent
\textbf{T4 clone detection}: a T4 code clone is also known as a semantic code clone. \textsf{AnDarwin}  \cite{AnDarwin} uses semantic information to find similar Android applications. \textsf{SrcClone}  \cite{alomari2020srcclone} uses program slicing technology to analyze code segment similarity to detect T4 clones. \textsf{SCDetector}  \cite{wu2020scdetector} combines the information about the token and the control flow graphs and uses a neural network to generate a code clone detector, achieving very good evaluation results. \textsf{MSCCD} does not have the ability to detect semantic clones and cannot be compared with these tools.

\noindent
\textbf{Big-Code clone detection}: \textsf{SourcererCC} \cite{Sajnani16} has increased the scalability of the code clone detection tool to 250 MLOC. Moreover, \textsf{SAGA}  \cite{li2020saga} detects code clones using a GPU accelerated suffix-array matching algorithm and raises the scalability to the level of 1 BLOC. Due to the amount of calculation, \textsf{MSCCD} is unable to outperform these tools on scalability.

In the era of \textbf{multilingual code clone detection}, \textsf{CCFinderSW} \cite{CCFinderSW} had the highest language extensibility before \textsf{MSCCD} was proposed. \textsf{CCFinderSW} generates a source-code parser by converting the grammatical rules of the target language into regular expressions. The parser can remove comments from the source code and generate a token sequence to match \textsf{CCFinderX} 's \cite{kamiya2021ccfinderx} clone detector. When the comment syntax of the target language cannot be converted into regular expressions, \textsf{CCFinderSW} cannot support it. Additionally, \textsf{CCFinderSW} only supports the detection of T2 clones. Both weaknesses have been resolved in \textsf{MSCCD}. Some text-based tools \cite{ducasse1999language} can also support multiple languages, but the similarity information available in such ancient technologies is too scant to detect more clone types. Furthermore, some existing tools claim to be easy to extend to new languages, but it turns out that the language extensibility of these tools is not enough.

Different from traditional detection tools, \textbf{Cross-language code clone detection} aims to detect code cloning between different languages. Perez et al. \cite{perez2019cross} detected clones between Java and Python by learning token-level vector representations and an LSTM-based neural network. \textsf{CLCDSA} \cite{CLCDSA_CLCCD} can detect cross-language clones without generating an intermediate representation by learning and comparing the similarity of features. \textsf{LICCA}  \cite{LICCA_CLCCD} extracts syntactic and semantic similarities based on the high-level representation of code from the SSQSA platform and can detect clones between five languages, including Java and C. \textsf{MSCCD} does not have the capability of cross-language clone detection. However, we expect that source code normalization with the same language extensibility as \textsf{MSCCD} can be used for cross-language clone detection. % 最後の文の意味が取れていません。but we expect that \textsf{MSCCD} 's parsing, and toke-bag extraction can be used in such cross-language code-clone detection. ???

There have been studies that target programs in which multiple languages are mixed  \cite{Moonen2001,Nikita2003,BACCHELLI2017}. 
In particular, web systems often contain programs in which multiple languages are mixed \cite{Rajapakse2007, Tariq2013}, and analysis tools for such systems are needed. Nakamura et al. also worked on code clone detection and suggested a detection tool for a web system  \cite{Nakamura2016}. Extending \textsf{MSCCD} to include such systems by using techniques such as island grammars \cite{Moonen2001,Nikita2003} is also a future challenge. Note that cross-language code clone detection assumes that each software system is written in a single language and should be distinguished from the case where multiple languages are mixed.

Determining how to benchmark code clone detection tools is a perennial problem for code clone researchers. Initially, applications to large-scale OSSs such as well-known operating systems (e.g., FreeBSD, Linux, and NetBSD) were frequently performed, and subsequent studies were compared by applying them to the same large-scale OSS as prior studies \cite{ccfinder,cp-miner,jiang2007deckard}. The first large-scale benchmark for a code clone detection tool was created by Bellon et al.  \cite{Bellon}. They visually judged whether a code clone was present. This benchmark is composed of C source code. The most prominent and recently used benchmark is BigCloneBench  \cite{BCB,BCE}, which was also used in this study. When creating this benchmark, Svajlenko and Roy succeeded in creating a larger benchmark than that of Bellon et al. by performing code mutations. This benchmark comprises Java source code.
In 2021, Svajlenko and Roy published a mutation and injection framework for benchmarking using mutation analysis  \cite{Svajlenko2021}.
There is also a growing body of research that uses competitive programming code, such as Google Code Jam, as benchmarks  \cite{DeepSim,perez2019cross,ASTNN}. 
Much of the research leveraging these benchmarks are deep-learning-based code clone detectors  \cite{DeepSim,perez2019cross,ASTNN}.
The Project CodeNet\footnote{\url{https://github.com/IBM/Project_CodeNet}}, which IBM recently released, will also be used as a benchmark in this research. Existing benchmarks are generally composed of Java or C source code \cite{BCB,Bellon}. With the increase in multilingual code clone detection, such as \textsf{MSCCD}, it is expected that benchmarks composed of source code in various languages will be created in the future.

\section{Conclusions and Future Work}
\label{sec8}

In this paper, we proposed \textsf{MSCCD}, a grammar-pluggable code clone detection tool. \textsf{MSCCD} showed the highest language scalability and good performance in terms of evaluation metrics, including recall and precision. In a case study with multiple languages, we discovered the impact of language features on code clone detection, revealing that further research in other languages is needed. 

In future works, we plan to evaluate \textsf{MSCCD} 's precision in various languages. We also intend to enable \textsf{MSCCD} to cover more clone types while maintaining the same language extensibility and further improving the scalability of \textsf{MSCCD}.

\begin{acks}
This work was supported by JST, PRESTO Grant Number JPMJPR21PA, Japan.
Also, this work was supported by JSPS KAKENHI Grant Numbers JP18H04094 and JP19K20240.
\end{acks}

\newpage

\bibliographystyle{ACM-Reference-Format}
\bibliography{MSCCD}

\end{document}